\documentclass[12pt]{article} 
\usepackage{latexsym}
\usepackage[dvipdfmx]{graphicx}
\usepackage{color}
\usepackage{ulem}

\begin{document} 
{ \pagestyle{empty} 
\vskip 3cm 
\centerline{\Large \bf Isotropic Metric in the Theory of General Relativity}
\vskip 10mm
\centerline{Kazuyasu Shigemoto \footnote{E-mail address:
shigemot@tezukayama-u.ac.jp}}
\centerline {{\it Department of Physics}}
\centerline {{\it Tezukayama University, Nara 631, Japan }}

\vspace{10mm}

\centerline{\Big[ {\large Tezukayama Academic Review, {\bf 3} (1997), 59-67} \Big]}

\vspace{20mm}

\centerline{\bf Abstract} \vspace{5mm} 
We explain why the isotropic metric is quite appropriate to put 
the physical meaning of spacial variables in the theory 
of general relativity.
Using the isotropic metric, we conclude that 
\ $i)$ $g_{00}$ does not become positive even inside the black hole, 
\ $ii)$ there exists the center of the Universe if the curvature 
of the Universe $k \ne 0$, 
\ $iii)$  the Universe is spacially finite but not colsed for $k>0$.
}

\vskip 0.4cm
\noindent
PACS number(s): 04.20.Jb, 04.60.-m, 95.30.Sf, 98.80.-k, 98.80.Hw
\hfil
\vfill
\newpage

\noindent {\large \bf \S 1.\ \ Introduction} 

\vspace{2mm} 

\indent
The theory of general relativity has a long history after Einstein's
work in 1916~\cite{Einstein}.  But even now, there remains some 
ambiguity to interprete the physical meaning of variables
in the metric.
The origin of this ambiguity comes from the general coordinate 
transformation in the theory of general relativity. 
The quite beautiful symmetry of the theory of general relativity,
the general coordinate invariance, makes the physical
meaning of variables ambiguous, because we can freely change variables.
Then the problem of time~\cite{Ours}, for example,
becomes the issue in the theory of general relativity.

In this paper, we will explain why the isotropic 
metric~\cite{Weiberg,Wheeler} is 
quite appropriate to put the physical meaning of spacial variables 
in the theory of general relativity.

Using the isotropic metric, we conclude that 
\ $i)$ $g_{00}$ does not become positive even inside the black hole, 
\ $ii)$ there exists the center of the Universe if the curvature 
of the Universe $k \ne 0$, 
\ $iii)$  the Universe is spacially finite but not colsed for $k>0$.

\vspace{2cm}

\noindent {\large \bf \S 2.  Physical Interpretation of the 
Isotropic Metric }

\vspace{2mm}

\qquad \qquad -- Isotropic Metric in Cartesian Coordinate --

\vspace{4mm}

We decompose the metric in the theory of general relativity into
the time and spacial metric ${d\sigma}^2$ in the form

\begin{eqnarray}
   &&ds^2= g_{\mu \nu} dx^{\mu} dx^{\nu}  
    = g_{00} dx^0 dx^0 +2 g_{0i} dx^{0} dx^{i} +{d\sigma}^2.
\label{e1}
\end{eqnarray}

In order to interprete the meaning of spacial variables 
physically, we use the Catresian corrdinate. 
We connect the Cartesian coordinate and 
the polar coordinate  just in the same 
way as in the flat space,  

\begin{eqnarray}
   && x=r \sin{\theta} \cos{\phi}, \quad y=r \sin{\theta} \sin{\phi}, 
   \quad z=r \cos{\theta}.  
\label{e3}
\end{eqnarray}

The polar coordinate in the theory of general relativity is 
ambiguous to put the physical meaning, because 
it is just one set of variables and nothing else in the curved 
space-time.  While, the Cartesian coordinate has the quite definite 
physical meaning.  Therefore we will always put
the physical meaning for spacial variables by transforming 
into the Cartesian coordinate. 
(It may be the issue whether the Cartesian coordinate only has the 
physical meaning or not, but at least the metric becomes uniquely fixed
by rewriting into the Cartesian coordinate.)

\vspace{1cm}

\noindent
{\underline {Isotropy of the space}

\vspace{2mm}

If the overal factor of the spacial metric is angle 
( $\theta$ and $\phi$ ) independent 
in comparison with the spacially flat metric in the polar coordinate, 
we call that such space is isotropic.
Isotropic spacial metric is therefore given by 

\begin{eqnarray}
  &&{d\sigma}^2= g_{i j} dx^{i} dx^{j} =f(r)(dr^2 +r^2 {d\theta}^2 
        +r^2 {\sin^2{\theta}}\ {d\phi}^2 ) \nonumber\\
  &&=f\left(\sqrt{x^2+y^2+z^2}\right)(dx^2+dy^2+dz^2).
\label{e4}
\end{eqnarray}

\vspace{1cm}

\noindent
{\underline{Uniquness of the spacial metric}}

\vspace{2mm}

We will explain how to uniquely determine physical spacial 
variables if the spacial metric has "isotropic" symmetry of the form

\begin{eqnarray}
  {d\sigma}^2=g(R) dR^2+h(R)(R^2 {d\Theta}^2 
+R^2  \sin^2{\Theta}\ {d\Phi}^2),
\label{e4'}
\end{eqnarray}

\noindent
where we assume that the asymptotic flatness,
$ g_{ij}(R,\Theta,\Phi)  \rightarrow \delta_{ij} $ as 
$R \rightarrow \infty$, is satisfied 
in order to interprete variables $R, \Theta, \Phi$ physically.

Then we can {\it uniquely}
determine the change of variables $ r=r(R), \Theta=\theta, \Phi=\phi$
in such a way as Eq.(\ref{e4'}) becomes in the form of Eq.(\ref{e4})
by keeping the asymptotic flatness $ g_{ij}(r,\theta,\phi)
 \rightarrow \delta_{ij} $ as $r \rightarrow \infty$. 

We can put the physical meaning of spacial variables 
in this isotropic metric, because we can easily transform the 
isotropic metric in polar coordinate into the isotropic metric in 
the Cartesian coordinate. Therefore there is no ambiguity of the 
physical interpretation of spacial variables,
such as the radius of the black hole.  

\vspace{1cm}

\noindent
{\underline{Mohogenuity of the space}

\vspace{2mm}

We call that the metric is homogeneous if the metric in Cartesian 
coordinate is invariant under the global parallel 
translation 

\begin{eqnarray}
   &&x \rightarrow x'=x+x_0 ,\nonumber\\
   &&y \rightarrow y'=y+y_0 ,\nonumber\\
   &&z \rightarrow z'=z+z_0 ,
\label{e5}
\end{eqnarray}

\noindent
where $x_0, y_0, z_0$ are constants.

Then the homogeneous spacial metric is given by 
 
\begin{eqnarray}
   {d\sigma}^2&&= g_{i j} dx^{i} dx^{j} ={\rm (const.)}(dr^2 +r^2 
                {d\theta}^2+r^2 {\sin^2{\theta}}\ {d\phi}^2 ) 
                   \nonumber\\  
              &&={\rm (const.)}(dx^2 + dy^2 + dz^2),
\label{e6}
\end{eqnarray}

\noindent
where ${\rm (const.)}$ is the $r, \theta, \phi$ or $x, y, z$ 
independent constant.

\vspace{2cm}

\noindent {\large \bf \S 3.  Isotropic Metric in the Black Hole}

\vspace{2mm}

The standard expression of Schwarzschild metric, which also express the 
metric of the black hole, is given by

\begin{eqnarray}
   ds^2&&= g_{\mu \nu} dx^{\mu} dx^{\nu}  \nonumber\\
       &&=-(1-2M/R){dt}^2 +\left( \frac{ {dR}^2 }{1-2M/R}
         +R^2 {d\Theta}^2 + R^2 \sin^2{\Theta}\ {d \Phi}^2 \right). 
\label{e6'}
\end{eqnarray}

We can connect this metric with the isotropic metric by the
relation 
$ R=r(1+M/2r)^2, \Theta=\theta, \Phi=\phi$ . Writing $R$ $=$ $
\displaystyle{ r\left(1+\frac{M}{r}+\frac{M^2}{4r^2} \right)}$ 
and noticing that $r=\sqrt{x^2+y^2+z^2}$ has a physical 
meaning so that $r$ must satisfy $r \ge 0$,
 we have $R=$ $\displaystyle{ M+r+\frac{M^2}{4 r} \ge 2 M} $, 
which means that 
\underline{  $g_{00} \le 0 $ even inside the black hole}. 
( If $R$ is allowed to become $R<2M$, then we may
interprete that the role of the space and time is exchanged for 
$d\Theta=0, d\Phi=0$, but this is not the case. )
Under the above change of variables, the spacial metric 
is given by\cite{Wheeler}

\begin{eqnarray}
   ds^2&&=-\left(\frac{1-M/2r}{1+M/2r} \right)^2 dt^2 
   +(1+M/2r)^4 
   (dr^2 +r^2 {d\theta}^2 +r^2 {\sin^2{\theta}}\ {d\phi}^2 )
    \nonumber\\   
    &&=-\left( \frac{1-M/2r}{1+M/2r} \right)^2 dt^2 
   +(1+M/2r)^4 (dx^2+dy^2+dz^2),  
\label{e7}
\end{eqnarray}

\noindent
where $r=\sqrt{x^2+y^2+z^2}$.  At $r=M/2$, which is equivalent $R=2r=M$,
we have $g_{00}=0$ but otherwise $ g_{00} >0$, so that even inside the 
black hole, the role of the space and time is {\it not} exchanged.

\vspace{2cm}

\noindent {\large \bf \S 4.\ \ Isotropic Metric in the Cosmology} 

\vspace{2mm} 

We apply the theory of general relativity to the cosmology,
and take the Friedmann Universe, and the metric\cite{Wheeler} of 
that Universe is given by 

\begin{eqnarray}
   &&ds^2= g_{\mu \nu} dx^{\mu} dx^{\nu}  \nonumber\\
   && =-dt^2 +a(t)^2 \left(\frac{dR^2}{1-k R^2}+R^2 {d\Theta}^2 
        +R^2 {\sin^2{\Theta}}\ {d\Phi}^2 \right) \nonumber\\
   && =-dt^2 +a(t)^2 \left(  \frac{dr^2+r^2 {d\theta}^2 
        +r^2 {\sin^2{\theta}}\ {d\phi}^2} 
             {(1+kr^2/4 )^2 }\right) \nonumber\\
   && =-dt^2 +a(t)^2 \left( \frac{dx^2+dy^2+dz^2}
       {\left[1+k(x^2+y^2+z^2)/4 \right]^2} \right),
\label{e8}
\end{eqnarray}

\noindent 
where 
\begin{eqnarray}
   && R= \frac{r}{(1+kr^2/4)},
   \quad \Theta=\theta, \quad \Phi=\phi, 
\label{e9} \\
   &&{\rm and\ }\quad  r=\sqrt{x^2+y^2+z^2}. 
\end{eqnarray}

From this expression, we can see that 
$\displaystyle{ 0 \le R=\frac{1}{(1/r+kr/4)} \le 1/\sqrt{k}}$ and 
$R=1/\sqrt{k}$ does not correspond to the infinity far point but
only finite distant point $ r=\sqrt{x^2+y^2+z^2}=2/\sqrt{k}$.
As $r$ changes as $ 0 \rightarrow 2/\sqrt{k} \rightarrow 
+\infty$, the corresponding $R$ changes as $0 \rightarrow 
1/\sqrt{k} \rightarrow 0$. Therefore $r=0$ and $r=\infty$ becomes
degenerate in $R$ variable. 
We also notice that 
\underline{there exists the center of the Universe for $k \ne 0$},
because the spacial metric has no invariance under the global parallel 
translation $x \rightarrow x+x_0, y \rightarrow y+y_0, 
z \rightarrow z+z_0$. 

\vspace{2cm}

\noindent
{\underline{ Finite Universe}}

\vspace{2mm}

We consider $k>0$ case, and first consider the total 
volume of the Universe, 

\begin{eqnarray}
    && {\rm V}= \int_{-\infty}^{+\infty} dx \int_{-\infty}^{+\infty} dy
         \int_{-\infty}^{+\infty} dz \sqrt{ \det{g_{ij}} }   \nonumber\\
    &&=\int \int \int_{-\infty}^{+\infty} 
       dx dy dz \frac{1}{\left[1+k(x^2+y^2+z^2)/4\right]^3}  
    \nonumber\\
    &&=\int_{0}^{+\infty} dr \int_{0}^{\pi} d\theta \int_{0}^{2 \pi}
    d\phi
    \frac{\ r^2 \sin{\theta} }{(1+kr^2/4)^3} \nonumber\\
    &&=\frac{16 \pi} {k^{3/2}} \int_{0}^{+\infty}d\xi 
     \frac{ \sqrt{\xi} }{ (1+\xi)^3} =\frac{2 \pi^2}{k^{3/2}}.
\label{e10} 
\end{eqnarray}

It may possible to calculate the total volume ${\rm V'}$
of the Universe with variables $R, \Theta, \Phi$ in the form

\begin{eqnarray}
    && {\rm V'}= \int_{0}^{\sqrt{k}} dR \int_{0}^{\pi} d\Theta 
    \int_{0}^{2 \pi} d\Phi
    \frac{\ R^2 \sin{\Theta} }{\sqrt{1-k R^2}} \nonumber\\
   &&= \frac{4 \pi}{k^{3/2}} \int_{0}^{1} dx \frac{x^2}{\sqrt{1-x^2}}
   =\frac{\pi^2}{k^{3/2}}.
\label{e10'}
\end{eqnarray}

\noindent 
We notice that ${\rm V}$ and ${\rm V'}$ gives the different value
and have the relation ${\rm V}=2{\rm V'}$. There is the reason of 
this factor $2$ difference.   For given $R$, there exist 
two $r_{+}, r_{-}$ with $ 0 \le r_{-} \le 2/\sqrt{k} \le r_{+}$,
which satisfy $ R=$ 
$\displaystyle{ \frac{1}{(1/r_{-}+kr_{-}/4)} }=$
 $\displaystyle{ \frac{1}{(1/r_{+}+kr_{+}/4)} }$, that is, 
the mapping between  
$r$ and $R$ is $2$ to $1$ except at $r=2/\sqrt{k}$.  
For example, $r=2/\sqrt{k}$ 
correspond to single $R=1/\sqrt{k}$, so that the correspondence is 
$1$ to $1$ only at this point. While $r=0$ and $r=+\infty$ both
corresponds to $R=0$. Thus the correct total volume 
of the Universe becomes ${\rm V}$ instead of ${\rm V'}$.  

Therefore we can see that the total volume of the Universe is finite. 
We notice that the Universe is {\it not} closed 
by looking at the expression of the spacial metric, which
is written by the Cartesian 
coordinate 

\begin{eqnarray}
   &&{d\sigma}^2=a(t)^2 \left( \frac{dx^2+dy^2+dz^2}
       {\left[1+k(x^2+y^2+z^2)/4 \right]^2} \right), 
\label{e11}
\end{eqnarray}

\noindent 
which gives 
$\displaystyle {g_{ij}(x,y,z,t)= a(t)^2 \delta_{ij} 
/\left[1+k(x^2+y^2+z^2)/4 \right]^2 }$.
From this expression, we can interprete that
\ $i)$ there is the center of the Universe for $k \ne 0$, because 
the spacial metric $g_{ij}$ is not symmetric 
under $ x \rightarrow x+x_0$,
$ y \rightarrow y+y_0$, $ z \rightarrow z+z_0$,
\ $ii)$ this spacial metric $g_{ij}$ is not periodic under 
$ x, y, z \rightarrow \pm \infty$.

Therefore we can conclude that 
\underline{ the Universe is {\it finite} but 
{\it not closed} for $k>0$ }. 

\vspace{1cm}

\noindent
\underline{Finite Universe in polar coordinate}

\vspace{2mm}

We will explain that the Universe is {\it finite} but {\it not closed} 
by using the polar coordinate also. 
We define $\sin{\chi}=\sqrt{k} R$, then we have 
$\cos{\chi}=\pm \sqrt{1-k R^2}$ but how should we 
choose the sign of $\pm$?
As $R$ changes from $0$ to $1/\sqrt{k}$ and cover that range only once, 
we must choose the range of $\chi$ as $ 0 \le \chi \le \pi/2$ and
we must choose $\cos{\chi}=\sqrt{1-k R^2}$ in order to
make that parametrization meaningful.
In this range of $\chi$, the mapping $ R \mapsto \chi$ and 
$ \chi \mapsto R$ both become single valued. 
By using the relation 
$\displaystyle{ R=\frac{1}{(1/r+kr/4)} }$, \ $i)$ $r=0$ gives 
$R=0$ and $\chi=0$ ,
\ $ii)$ $r=2/\sqrt{k}$ gives $R=1/\sqrt{k}$ and $\chi=\pi/2$, 
\ $iii)$ $r=+\infty$ gives $R=0$ and $\chi=0$. 
We will comapare this parametrization of spacial metric with the 
parametrization of the half ${\rm S^3}$. 

\vspace{2mm}

\noindent
\underline{Parametrization of the half ${\rm S^3}$}

We can parametrize the half ${\rm S^3}$ in the form 

\begin{eqnarray}
   &&w_1=\frac{ \sin{\chi} \sin{\Theta} \cos{\Phi} }{\sqrt{k}},  
   \nonumber\\
   &&w_2=\frac{ \sin{\chi} \sin{\Theta} \sin{\Phi} }{\sqrt{k}},  
   \nonumber\\
   &&w_3=\frac{ \sin{\chi} \cos{\Theta} }{ \sqrt{k} }, \nonumber\\
   &&w_4=\frac{ \cos{\chi} }{ \sqrt{k} } ,
\label{e12}
\end{eqnarray}

\noindent 
with 
\begin{eqnarray}
   && 0 \le \chi \le \pi/2, \quad   0 \le \Theta \le \pi \quad
   0 \le \Phi \le 2 \pi. 
\label{e13}
\end{eqnarray}

\noindent
We call the above parametrization as the half ${\rm S^3}$, because 
the range of $\chi$ in closed ${\rm S^3}$ is 
$0 \le \chi \le \pi$, while $\chi$ in Eq.({\ref{e13}) 
takes only the half compared to ${\rm S^3}$. Then we can 
understand that the above half ${\rm S^3}$ is not closed.

In these variables, we have 
${w_{1}}^2+{w_{2}}^2+{w_{3}}^2+{w_{4}}^2=1/k$
and the spacial metric is given by

\begin{eqnarray}
  && {d\sigma}^2=a(t)^2 \left({dw_{1}}^2+{dw_{2}}^2+{dw_{3}}^2
     +{dw_{4}}^2\right) \nonumber\\
  &&=a(t)^2 \left( \frac{dR^2}{1-k R^2}
     +R^2 {d\Theta}^2 +R^2 {\sin{\Theta}}^2 {d\Phi}^2\right).  
\label{e13'}
\end{eqnarray}

We will explain some interesting paths in the followings.

\vspace{5mm}

\noindent
\underline{path(1)} 

\vspace{2mm}

We consider a path, going around the circle at finite distance
 $x^2+y^2=4/k$ with $z=0$ in Cartesian coordinate, 
which is equivalent to the path 
changing $\phi$ from $0$ to $2 \pi$ with fixed $r=2/\sqrt{k}$, 
$\theta=\pi/2$ in polar coordinate. As before, we connect the 
Cartesian coordinate and the polar coordinate in the form
 $x=r \sin{\theta} \cos{\phi}$, $y=r \sin{\theta} \sin{\phi}$, 
 $z=r \cos{\theta}$. 

This path is equivalent to the path 
changing $\Phi$  from $0$ to $2 \pi$ with fixed
$\chi=\pi/2$ ( $R=1/\sqrt{k}$ ), $\Theta=\pi/2$ in 
$\chi, \Theta, \Phi$ variables.
This path is the great circle even for the non-closed 
half ${\rm S^3}$,
that is, though $\chi$ changes in the range 
$ 0 \le \chi \le \pi/2$, the above path becomes the great circle.
Of course this path is the great circle 
on the closed ${\rm S^3}$ also.

\vspace{5mm}

\noindent
\underline{path(2)}

\vspace{2mm} 

What we are interested in is the path 

\begin{eqnarray}
   &&x: 0 \rightarrow 2/\sqrt{k} \rightarrow +\infty  
   \rightarrow 2/\sqrt{k} \rightarrow 0 
   \rightarrow -2/\sqrt{k} \rightarrow -\infty \rightarrow 
    -2/\sqrt{k} \rightarrow 0, \nonumber\\
   &&y=z=0.   \nonumber
\end{eqnarray}

In polar coordinate, this path is expressed as the result
of the following steps:

\begin{eqnarray}
{\rm Step\ 1)}&&  \nonumber \\
  &&r: 0 \rightarrow 2/\sqrt{k} \rightarrow +\infty 
  \rightarrow 2/\sqrt{k} \rightarrow 0,  \nonumber\\
  && \theta=\pi/2, \quad \phi=0,       \nonumber\\
{\rm Step\ 2)}&&  \nonumber \\
  &&\phi: 0 \rightarrow \pi, \nonumber\\
  && r=0, \quad \theta=\pi/2,       \nonumber\\
{\rm Step\ 3)}&&  \nonumber \\
  &&r: 0 \rightarrow 2/\sqrt{k} \rightarrow +\infty 
  \rightarrow 2/\sqrt{k} \rightarrow 0,  \nonumber\\
  && \theta=\pi/2, \quad \phi=\pi.       \nonumber
\end{eqnarray}

Using $\chi, \Theta, \Phi$ variables on 
the non-closed half ${\rm S^3}$,
this path is given by 

\begin{eqnarray}
{\rm Step\ 1)}&&  \nonumber \\
  &&\chi: 0 \rightarrow \pi/2 \rightarrow 0 
  \rightarrow \pi/2 \rightarrow 0,  \nonumber\\
  && \Theta=\pi/2, \quad \Phi=0,       \nonumber\\
{\rm Step\ 2)}&&  \nonumber \\
  &&\Phi: 0 \rightarrow \pi, \nonumber\\
  && \chi=0, \quad \Theta=\pi/2,       \nonumber\\
{\rm Step\ 3)}&&  \nonumber \\
  &&\chi: 0 \rightarrow \pi/2 \rightarrow 0 
  \rightarrow \pi/2 \rightarrow 0,  \nonumber\\
  && \Theta=\pi/2, \quad \Phi=\pi,       \nonumber
\end{eqnarray}

\noindent
and this path is {\it not} the great circle, but the path going to the 
edge ($\chi=\pi/2$) and coming back to the origin. 
In this way, the above path is not
the closed path on the closed manifold, but the path going outside 
and coming back to the origin on the non-closed half ${\rm S^3}$.

While on the closed ${\rm S^3}$, where $\chi$ takes 
$ 0 \le \chi \le \pi$, we can take the path
\begin{eqnarray}
{\rm Step\ 1)}&&  \nonumber \\
  &&\chi: 0 \rightarrow \pi/2 \rightarrow \pi 
  \rightarrow \pi/2 \rightarrow 0  \nonumber\\
  && \Theta=\pi/2, \quad \Phi=0       \nonumber\\
{\rm Step\ 2)}&&  \nonumber \\
  &&\Phi: 0 \rightarrow \pi \nonumber\\
  && \chi=0, \quad \Theta=\pi/2       \nonumber\\
{\rm Step\ 3)}&&  \nonumber \\
  &&\chi: 0 \rightarrow \pi/2 \rightarrow \pi 
  \rightarrow \pi/2 \rightarrow 0  \nonumber\\
  && \Theta=\pi/2, \quad \Phi=\pi       \nonumber
\end{eqnarray}

\noindent 
and this is the great circle on the closed ${\rm S^3}$.

\vspace{20mm}

\noindent {\large \bf \S 5. Summary }

\vspace{2mm}

In this paper, we interprete the meaning of 
spacial variables physically by rewriting into the Cartesian coordinate. 
In this context, the isotropic metric is quite 
appropriate to put the physical meaning of spacial variables 
in the theory of general relativity.  The metric also becomes unique 
after rewriting into the Cartesian coordinate. 
Using the isotropic metric, we conclude that 
\ $i)$ $g_{00}$ does not become positive even inside the black hole, 
\ $ii)$ there exists the center of the Universe if the curvature 
of the Universe $k \ne 0$, 
\ $iii)$  the Universe is spacially finite but not colsed for $k>0$.

\vspace{20mm}

\noindent {\large \bf Acknowledgements}

\vspace{3mm}

\indent
Author is grateful to the Special Research Fund 
at Tezukayama Univ. for financial support. 

\vspace{10mm}



\end{document}